\documentclass[10pt,rmp,twocolumn,showpacs,floats,letterpaper]{revtex4}

\usepackage{times}
\usepackage{mathptmx}
\usepackage{graphicx}
\usepackage{dcolumn}
\usepackage{bm}

\bibpunct{}{}{,}{s}{}{}

\begin{document}

\title{An ion-implanted silicon single-electron transistor}

\author{V.C. Chan}
 \email{victor.chan@bigpond.com}
\author{D.R. McCamey}
\author{T.M. Buehler}
 \altaffiliation[Now at ]{ABB Switzerland, Corporate Research.}
\author{A.J. Ferguson}
\author{D.J. Reilly}
 \altaffiliation[Now at ]{Dept. Physics, Harvard University, Cambridge, 02138, USA.}
\author{A.S. Dzurak}
\author{R.G. Clark}

\affiliation{
Centre for Quantum Computer Technology, Schools of
Physics and Electrical Engineering \& Telecommunications,
The University of New South Wales, NSW 2052, Sydney Australia\\
}

\author{C. Yang}
\author{D.N. Jamieson}

\affiliation{ Centre for Quantum Computer Technology, School of
Physics, University of Melbourne, VIC 3010, Australia\\
}

\date{\today}

\begin{abstract}
We report on the fabrication and electrical characterization at
millikelvin temperatures of a novel silicon single-electron
transistor (Si-SET). The island and source-drain leads of the
Si-SET are formed by the implantation of phosphorus ions to a
density above the metal-insulator-transition, with the tunnel
junctions created by undoped regions. Surface gates above each of
the tunnel junctions independently control the tunnel coupling
between the Si-SET island and leads. The device shows periodic
Coulomb blockade with a charging energy e$^2$/2C$_\Sigma$ $\sim$
250 $\mu$eV, and demonstrates a reproducible and controllable
pathway to a silicon-based SET using CMOS processing techniques.
\end{abstract}


\maketitle

Single-electron transistors (SETs) have excellent potential as
elementary devices in large scale circuits due to their small
dimensions and low power dissipation. Applications for SETs in
single-electron logic and memory cells \cite{heij,stone} have also
been demonstrated. Silicon-based SETs (Si-SETs) are of particular
interest because of their compatibility with CMOS technology.
While the fabrication of Si-SETs devices is a non-trivial process,
a number of different approaches have been demonstrated such as
gated two-dimensional electron gases (2DEGs) \cite{khoury},
pattern dependent etching of silicon-on-insulator (SOI) material
\cite{ntt,johnnygoreman}, and random formation of Coulomb islands
by nano-particles or defects in silicon
\cite{specht,baron,freddy,anri,kondo}. Some difficulties of
integrating such approaches into more complex single-electron
circuits include inconsistent island formation resulting in random
electron potentials and excessive gating requirements for
electrostatically defining Coulomb islands and tunnel junctions.

\begin{figure}
\includegraphics[width=7cm]{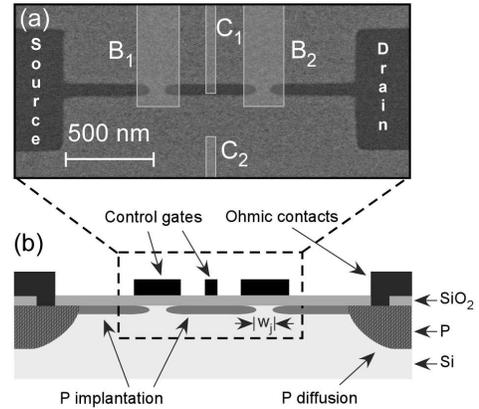} \caption{\label{fig:one}
(a) Scanning electron micrograph of a device prior to RTA. The
darker areas indicate regions of phosphorus ion implantation
damage. Surface gates are shown schematically. (b) Cross-section
schematic of the device showing the phosphorus implanted area,
phosphorus diffusion, ohmic contacts and surface gates.}
\end{figure}

The work presented in this letter features a Si-SET comprising of
a single well-defined island and source-drain leads fabricated by
a patterned phosphorus ion implantation process. This approach is
\emph{not reliant on random island formation}, and offers a new
method for \textit{controlled, reliable and reproducible}
formation of Coulomb islands. This Si-SET essentially consists of
two nano-scale metal-oxide-semiconductor field effect transistors
(MOSFETs) in series with a micron scale central island, with the
MOSFETs acting as tuneable tunnel junctions. As the island is
solely defined by the patterned ion implantation, no gates are
required to electrostatically define the island structure.
Furthermore, the controlled formation of well-defined islands
makes the coupling of multiple single-electron devices for more
complex circuits a simpler task. We discuss the device fabrication
and electrical measurements performed at T = 50 mK in which
Coulomb blockade behavior is observed.

Devices were fabricated on a high resistivity ($>$ 5 k$\Omega$.cm)
n-type silicon wafer. Firstly, ohmic contacts for the source and
drain leads of the device were defined via phosphorus diffusion. A
5 nm gate oxide was then grown using a thermal oxidation process.
Electrical characterization of (large-scale) MOSFET devices
fabricated with gate oxides grown using the same process indicate
typical trap densities of $2 \times 10^{11}$ cm$^{-2}$ at T = 4.2
K \cite{DRM}. High-resolution TiPt (15 nm Ti, 65 nm Pt) alignment
markers, 100 nm $\times$ 100 nm in dimension, were defined by
electron-beam lithography. These markers were used to align
subsequent e-beam lithography steps with an accuracy of $\leq$ 50
nm, whilst being able to withstand subsequent thermal processes. A
150 nm thick poly-methyl-methacrylate (PMMA) resist was applied
and patterned by e-beam lithography for use as a mask for ion
implantation. The dimensions of the SET island were designed to be
70 nm $\times$ 500 nm. A number of different devices were
fabricated with intrinsic tunnel junction widths w$_{\textrm{j}}$
of 100 nm and 150 nm (see figure \ref{fig:one}(b)).

Phosphorus ions at 14 keV were implanted with an areal dose of
approximately $1.22$ $\times$ $10^{14}$ cm$^{-2}$ and a mean
implantation depth of $\sim$ 20 nm. This equates to approximately
43000 ions in the island and a doping density of $\sim 10^{19}$
cm$^{-3}$. A scanning electron micrograph (SEM) of a device after
ion implantation and removal of the PMMA mask is shown in figure
\ref{fig:one}, note the high contrast between the ion implanted
and masked regions due to damage in the silicon crystal caused by
the ion implantation. A rapid thermal anneal (RTA) at $1000^{o}$C
for 5 sec was performed to repair implantation damage and
electrically activate the phosphorus donors. Post-RTA SEM imaging
of the implanted regions show significantly less contrast to the
masked regions, indicating repair of the implantation damage. In
the final step, four electrostatic control gates (shown
schematically in figure \ref{fig:one}) were fabricated by e-beam
lithography using 60 nm PMMA resist and TiAu (10 nm Ti, 20 nm Au)
metallization. Two barrier gates (B$_{\textrm{1}}$ and
B$_{\textrm{2}}$) were defined over the tunnel junctions, one gate
(C$_{\textrm{1}}$) over the island, and the fourth gate
(C$_{\textrm{2}}$) a small distance away from the island.

\begin{figure}
\includegraphics[width=7cm]{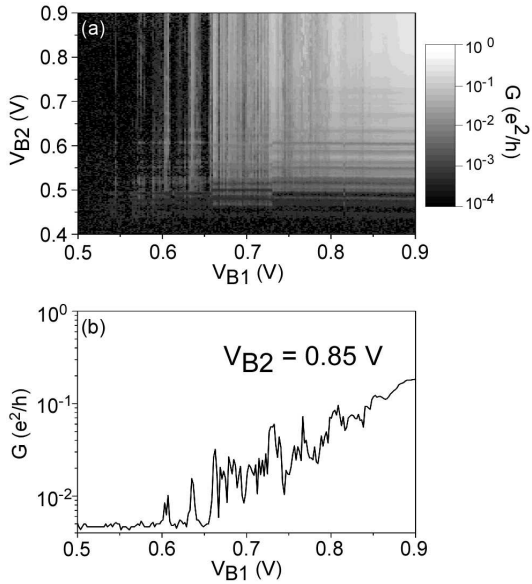}
\caption{\label{fig:two} (a) Si-SET (w$_{\textrm{j}}$ = 150 nm)
conductance intensity plot as a function of V$_{\textrm{B1}}$ and
V$_{\textrm{B2}}$. Resonances, most likely due to defects in the
tunnel junctions, are observed in the vertical and horizontal axis
indicating independent control of the tunnel barriers. (b) Single
trace taken from (a) at V$_{\textrm{B2}}$ = 0.85 V showing
increasing conductance with positive voltage applied to
B$_{\textrm{1}}$. V$_{\textrm{ac}}$ = 100 $\mu$V. T = 50 mK.}
\end{figure}

Electrical measurements were carried out at T = 50 mK in a
dilution refrigerator using standard lock-in techniques at
frequencies $<$ 200 Hz. Initial characterization of these devices
focussed on the effect of the electrostatic barrier gates on the
tunnel junctions. Figure \ref{fig:two}(a) shows the conductance of
a device with 150 nm wide tunnel junctions as a function of
V$_{\textrm{B1}}$ and V$_{\textrm{B2}}$. The behavior observed in
this device is representative of all of the devices that were
measured. With increasing barrier gate voltage, the device
conductance is increased as expected for an enhancement-mode
MOSFET (see figure \ref{fig:two}(b)). Resonances observed in the
device conductance are visible in figure \ref{fig:two}(a) as
vertical and horizontal lines, and indicate that each barrier gate
is primarily coupled to its respective tunnel junction. These
resonances most likely arise from the random potentials in the
tunnel junctions, due to stray dopants, charge traps and other
defects. These form unintentional islands that exhibit either
Coulomb blockade or resonant tunnelling phenomena. \cite{sanq}.
Even with these resonances present, the measurements indicate good
gate control of the overall conductance of the tunnel junctions.

\begin{figure}
\includegraphics[width=7cm]{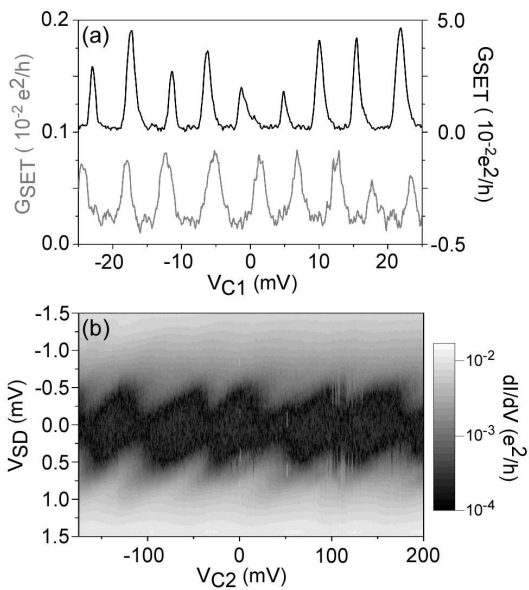}
\caption{\label{fig:three} (a) Si-SET (w$_{\textrm{j}}$ = 100 nm)
source-drain conductance as a function of V$_{\textrm{C1}}$ shows
Coulomb blockade oscillations for V$_{\textrm{B1}}$ =
V$_{\textrm{B2}}$ = 0 mV (grey trace, V$_{\textrm{SD}}$ =
350$\mu$V) and V$_{\textrm{B1}}$ = V$_{\textrm{B2}}$ = -200 mV
(black trace, V$_{\textrm{SD}}$ = 0$\mu$V). V$_{\textrm{ac}}$ =
50$\mu$V. T = 50 mK. (b)  Si-SET (w$_{\textrm{j}}$ = 100 nm) bias
spectroscopy measurement. Coulomb blockade diamonds are observed
indicating a device charging energy e$^2$/2C$_\Sigma$ of $\sim$
250 $\mu$eV. Around V$_{\textrm{C2}}$ = 110 $\mu$V the device
conductance is perturbed by charge noise fluctuation.
V$_{\textrm{ac}}$ = 50$\mu$V. T = 50 mK.}
\end{figure}

In figure \ref{fig:three}(a), the grey trace shows the conductance
of a Si-SET with 100 nm wide tunnel junctions as a function of
V$_{\textrm{C1}}$ with V$_{\textrm{B1}}$ = V$_{\textrm{B2}}$ = 0
mV and V$_{\textrm{SD}}$ = 350 $\mu$V. In the measurement,
periodic Coulomb blockade oscillations are observed indicating a
constant capacitance between the gate and the Si-SET island. The
consistent form of the conductance peaks also indicates that the
tunnel junctions are not significantly changing. Figure
\ref{fig:three}(b) shows a bias spectroscopy measurement for the
same device. Coulomb charging diamonds with a constant charging
energy are observed, consistent with charging a single metallic
island as opposed to a random minimum potential in the tunnel
junctions. This is in contrast to Coulomb blockade associated with
unintentionally formed islands in the channel of FETs where the
charging energy and periodicity changes significantly with island
occupancy. In common with other SETs, the device is sensitive to
1$/$f charge noise and nearby two-level fluctuators (TLF) which
perturb the device conductance, as observed in figure
\ref{fig:three} (b) around V$_{\textrm{C2}}$ = 110 $\mu$V.

Further measurements were performed on the Si-SET with 100 nm wide
tunnel junctions to observe how gate tuning of the tunnel barriers
could be used to control the device conductance whilst maintaining
Coulomb blockade behavior. A number of different voltages were
applied to the barrier gates (B$_{\textrm{1}}$ and
B$_{\textrm{2}}$). Under low source-drain bias conditions and
V$_{\textrm{B1}}$ = V$_{\textrm{B2}}$ = 0 V, the device
conductance is actually below the measurable threshold for
standard lock-in amplifier techniques (I$_{\textrm{SD}}$ $<$ 1
pA). At V$_{\textrm{B1}}$ = V$_{\textrm{B2}}$ = - 200 mV, the
barrier gates are biased close to some resonance in the barriers
which results in an overall increase in the conductance of the
device. The black trace in figure \ref{fig:three}(a) shows the
Si-SET conductance under these barrier conditions and Coulomb
blockade oscillations are clearly observed. In comparison to the
grey trace in the same figure, which shows Si-SET conductance when
V$_{\textrm{B1}}$ = V$_{\textrm{B2}}$ = 0 mV and V$_{\textrm{SD}}$
= 350 $\mu$V, the peak conductance observed in the black trace is
an order of magnitude higher (G$_{\textrm{SET}}$ (grey) $\sim$ 7
$\times$ 10$^{\textrm{-3}}$ e$^{\textrm{2}}$/h and
G$_{\textrm{SET}}$ (black) $\sim$ 4 $\times$ 10$^{\textrm{-2}}$
e$^{\textrm{2}}$/h). The periodicity of the oscillations are
consistent between both traces reinforcing the notion that Coulomb
blockade in a single well-defined metallic island is being
observed. Similar behavior is seen for barrier conditions where
large positive voltages are applied to the barrier gates to
increase the transparency of the tunnel junctions.

From the bias spectroscopy data shown in figure
\ref{fig:three}(b), the charging energy of the island in the
Si-SET with 100 nm wide tunnel junctions is determined to be
e$^2$/2C$_\Sigma$ $\sim$ 250 $\mu$eV with a total capacitance
C$_\Sigma$ of 320 aF. The capacitances of gates C$_{\textrm{1}}$
and C$_{\textrm{2}}$ to the island are determined from the period
of the measured Coulomb blockade oscillations to be 27 aF and 2.1
aF respectively. The capacitances of barrier gates
B$_{\textrm{1}}$ and B$_{\textrm{2}}$ to the island were not
measured for this device, however these values have been measured
for other devices and are typically of order 100 aF. The asymmetry
of the Coulomb diamonds is a result of the different capacitances
between the implanted island and the source and drain leads, which
are determined from the Coulomb diamonds to be 21 aF and 55 aF.
This difference in capacitance can be attributed to the barriers
being of slightly different dimensions and differences in their
complex potential landscape.

A proof-of-principle has been demonstrated for the controlled
formation of Coulomb blockade islands using CMOS processing
techniques. The Si-SET demonstrates highly controllable
single-island charging behavior due to the well-defined electron
potential. The tunnel coupling between the Coulomb blockade island
and leads can be changed by using electrostatic gates above the
tunnel junctions (MOSFETs). The development of Si-SET as elements
in more complex single-electron circuits requires the ability to
controllably couple multiple islands and future work will focus on
this. Whilst characterization of this device demonstrates Si-SET
behavior at low temperature (T = 50 mK), further development of
this fabrication technique will involve scaling down of the island
size to enable the charging energy E$_{\textrm{C}}$ to dominate at
higher temperatures (E$_{\textrm{C}}$ $>$ k$_{\textrm{B}}$T). In
addition, related devices may provide a platform for the study of
electron transport between locally doped regions in silicon, which
is of particular relevance to silicon-based quantum computing
\cite{kane,HOLL,schenk}.

The authors would like to thank S.E. Andresen, J.C. McCallum, M.
Lay, C.I. Pakes, and S. Prawer for helpful discussions and E.
Gauja, R. P. Starrett, D. Barber, G. Tamanyan and R. Szymanski for
their technical support. This work was supported by the Australian
Research Council, the Australian Government and by the US National
Security Agency (NSA), Advanced Research and Development Activity
(ARDA) and the Army Research Office (ARO) under contract number
DAAD19-01-1-0653.

\bibliography{si-set}
\end{document}